%% file: ms3992.tex
\documentclass{aa}
\usepackage{graphics}
\usepackage{natbib}

\begin{document}
\def\ab{$\sim$}
\def\frac{$''$\hspace*{-.1cm}}
\def\deg{$^{\circ}$\hspace*{-.1cm}}
\def\min{$'$\hspace*{-.1cm}}
\def\h2{H\,{\sc ii}}
\def\hi{H\,{\sc i}}
\def\hb{H$\beta$}
\def\ha{H$\alpha$}
\def\hd{H$\delta$}
\def\heii{He\,{\sc ii}}
\def\hg{H$\gamma$}
\def\sii{[S\,{\sc ii}]}
\def\siii{[S\,{\sc iii}]}
\def\oiii{[O\,{\sc iii}]}
\def\oii{[O\,{\sc ii}]}
\def\hei{He\,{\sc i}}
\def\sm{$M_{\odot}$}
\def\slum{$L_{\odot}$}
\def\mdot{$\dot{M}$}
\def\x{$\times$}
\def\sec{s$^{-1}$}
\def\cm2{cm$^{-2}$}
\def\mcube{$^{-3}$}
\def\lam{$\lambda$}

\def\ac{Al~{\sc{iii}}\ }
\def\cb{C\,{\sc{ii}}}
\def\cc{C\,{\sc{iii}}}
\def\cd{C\,{\sc{iv}}}
\def\ca{Ca\,{\sc{ii}}}
\def\crb{Cr\,{\sc{ii}}}
\def\feb{Fe\,{\sc{ii}}}
\def\fec{Fe\,{\sc{iii}}}
\def\hea{He\,{\sc{i}}}
\def\heb{He\,{{\sc ii}}}
\def\nb{N\,{\sc{ii}}}
\def\nc{N\,{\sc{iii}}}
\def\nd{N\,{\sc{iv}}}
\def\ne{N\,{\sc{v}}}
\def\na{Na~{\sc{i}}\ }
\def\nf{Ne~{\sc{i}}\ }
\def\soc{S~{\sc{iii}}\ }
\def\sd{S\,{\sc{iv}}}
\def\sib{Si\,{\sc{ii}}}
\def\sic{Si\,{\sc{iii}}}
\def\sid{Si\,{\sc{iv}}}
\def\tib{Ti\,{\sc{ii}}}

\title{The stellar environment of SMC N81\thanks
   {Based on observations obtained at the European Southern 
   Observatory, Paranal, Chile;  Program 69.A-0123(A)}\,$^{_{'}}$\,\thanks{
Based on observations made with the NASA/ESA Hubble Space Telescope,
   obtained at the Space Telescope Science
   Institute, which is operated by the Association of Universities for
   Research in Astronomy, Inc., under NASA contract NAS 5-26555. These
   observations are associated with program \# 6535.}
}

\offprints{Fr\'ed\'eric Meynadier, \hspace{1cm} \\Frederic.Meynadier@obspm.fr}

\date{Received 15 May 2003/ Accepted 3 September 2003}

\titlerunning{SMC N81 
}
\authorrunning{Heydari-Malayeri et al.}

\author{M. Heydari-Malayeri\inst{1} \and F. Meynadier\inst{1} 
\and V. Charmandaris\inst{2,1} \and L. Deharveng\inst{3} \and 
Th. Le Bertre\inst{1} \and 
M.R. Rosa \inst{4,}\,\thanks{
    Affiliated to the Astrophysics Division, Space Science Department of
    the European Space Agency.}
 \and D. Schaerer\inst{5,6} 
}

\institute{{\sc lerma}, Observatoire de Paris, 61 Avenue de l'Observatoire, 
F-75014 Paris, France \and Cornell University, Astronomy Department,
106 Space Sciences Bldg., Ithaca, NY 14853, U.S.A. \and Observatoire
de Marseille, 2 Place Le Verrier, F-13248 Marseille Cedex 4, France
\and Space Telescope European Coordinating Facility, European Southern
Observatory, Karl-Schwarzschild-Strasse-2, D-85748 Garching bei
M\"unchen, Germany  \and Observatoire de Gen\`eve, 51, Ch. des Maillettes, 
CH-1290 Sauverny,  Switzerland      
\and  Laboratoire d'Astrophysique, UMR 5572, Observatoire
        Midi-Pyr\'en\'ees, 14,  Avenue E. Belin, F-31400 Toulouse, France
}

\abstract{We present near infrared {\it JHK} imaging of the Small 
Magellanic Cloud compact \h2 region N81 using the ISAAC camera at the
ESO Very Large Telescope (Antu).  Our analysis of the stellar
environment of this young massive star region reveals the presence of
three new stellar populations in the surrounding field which are
mainly composed of low mass stars.  The main population is best fitted
by evolutionary models for \ab\,2\,\sm\, stars with an age of 1 Gyr.
We argue that these populations are not physically associated with the
\h2 region N81. Instead they are the result of a number of low mass
star forming events through the depth of the SMC south of its
Shapley's wing. The populations can rather easily be probed due to the
low interstellar extinction in that direction.  \\
}

\maketitle

\section{Introduction}

High-Excitation Blobs (HEBs) represent a rare class of compact \h2
regions in the Magellanic Clouds \citep{mhm82}.  In contrast to the
typical \h2 regions of these neighboring galaxies, which are extended
structures spanning several minutes of arc on the sky (more than 50
pc) and powered by a large number of hot stars, HEBs are dense small
regions usually 5\frac\, to 10\frac\, in diameter (1 to 3
pc). Moreover, they happen to lie adjacent or in the direction of the
typical giant \h2 regions, with the exception of SMC N81
\citep{henize} which has apparently been formed in isolation.  
They are probably the optical counterparts of the Galactic
ultracompact \h2 regions \citep{churchwell}  lying relatively close to
the cloud surface because the molecular cloud layers above the newborn
stars have been scraped by the strong UV field of the adjacent massive
stars or by the champagne flows \citep{tenorio79}. \\

Optical observations of a number of these objects, LMC N159-5, N160A1,
N160A2, N83B, N11A, as well as SMC N88A and N81, allowed to derive
their global physical characteristics and establish them as a
particular class of metal-poor
\h2\, regions in the Magellanic Clouds \citep{mhm82, 
mhm83, mhm85, mhm86, mhm90, testor85, mhm88}. 
In particular, it was shown that HEBs are generally very
affected by local dust (references above; see also \citet{israel91}).  
However, those early studies made it also clear that in order
to better understand the stellar properties in those small regions
sub-arcsecond spatial resolution was needed. \\

This was achieved by our {\it HST} WFPC2 high resolution imaging of
seven HEBs, SMC N81, N88A, LMC N159-5, N83B, N11A, N160A1, N160A2
\citep{mhm99a, mhm99b, mhm99c, mhm01a, mhm01b, mhm02a, mhm02b}. 
We were able to spatially resolve these objects for the first time,
uncovering their morphology, nebular features, the location of their
high excitation
\oiii\,$\lambda$5007 zones, and the variation of the extinction across
them. Those observations showed that powerful stellar winds and shocks
create magnificent scenes testifying to high activity: outstanding
emission ridges, cavities sculpted in the ionized gas, prominent dust
structures protruding from hot gas, and even unknown compact \h2 blobs
immersed in the HEBs and harboring very young hot stars. In a few
cases our {\it HST} images revealed some of the exciting stars, but
not all of them, and in some other cases we could not conclusively
identify the exciting stars at all. This is likely due to the high
dust content of those regions, which is rather remarkable given their
low metallicity \citep{bouchet85}.  \\

The \h2 region N81 lies well outside the main body of the SMC in 
the outer parts of Shapley's wing. Discovered by  \citet{shapley} 
as a ``large cloud of faint stars extending eastward from the SMC to
the LMC'', the wing was shown to be in fact the tail of  
a much larger \hi\, structure  linking the SMC to the LMC
\citep{kerr54, hindman63, mathewson84}. 
Models and observations suggest that 
the neutral hydrogen structures known as the Magellanic Bridge, the
Magellanic Stream, and the Leading Arm result from the Clouds'
interaction with each other and the Milky Way
\citep{murai80, moore94,gardiner96, putman98}.  
Several works support the finding that the SMC wing is pointing
towards the LMC, and is therefore closer to us than the SMC bar 
\citep{mathewson86, caldwell86, irwin90}.  Regarding N81 itself, which
lies towards the outer parts and south of the wing, our {\it HST}
observations showed the presence of a tight cluster of newborn massive
stars embedded in this nebular ``blob'' of
\ab\,10\frac\, across (\citet{mhm99a}, hereafter Paper I). Six of them
are grouped in the core region of \ab\,2\frac\, diameter, with two
of the main exciting stars, in the very center, separated by only
0\frac.27 or 0.08 pc. The images display violent phenomena such as
stellar winds, shocks, ionization fronts, typical of turbulent
starburst regions.  Follow-up far UV spectroscopy with STIS
(\cite{mhm02b}, Paper II) also revealed a particularly interesting case
in SMC N81, where several of the stars  are O6--O8
types, but display extremely weak wind profiles \citep{mart}.
The astonishing weakness of their wind profiles and the
sub-luminosity (up to \ab\,2 mag fainter in $M_{V}$ than the corresponding
dwarfs) make these stars a unique stellar population in the Magellanic
Clouds. Our analysis suggests that they are probably in the
Hertzsprung-Russell diagram locus of a particularly young class of
massive stars, the so-called Vz luminosity class, as they are arriving
at the zero age main sequence. \\

The heavy dust content of the HEBs calls for doing high spatial
resolution observations in the near infrared.  Our  aim in using
the VLT/ISAAC was to detect and study all massive embedded stars
inside N81, as well as the associated surrounding population.
However, due to the limited spatial resolution of the present 
observations, this paper is mainly focused on the properties of the stellar
environment of N81.\\

\section{Observations and data reduction}

The N81 region was observed in service mode with the ESO Very Large
Telescope (VLT). The infrared spectro-imager ISAAC was used at the
Nasmyth B focus of Antu through filters {\it Ks} on 7 October and $J$ and
$H$ on 18 November 2001. The infrared detector  (Hawaii Rockwell array) had
$1024\times1024$ pixels of 18.5$\mu$m each (0\frac.148 on the sky),
thus providing a field of 2\min.5\,\x\,2\min.5.
The seeing varied between 1\frac.02 and 1\frac.40\, ({\sc
fwhm}). \\

A set of individual, 10-second exposures were obtained in each filter
using a dithering method with a random offset of 15\frac\, at
most. The number of exposures were 10, 20, and 18 for the $J$, $H$,
and {\it Ks} bands respectively.  The data were processed using the
VLT pipeline and we verified that their quality was sufficient for our
intended science. One area of concern was that small fluctuations of
the sky near crowded areas of each frame could be due to sky
subtraction by the pipeline ``jitter'' recipe. Therefore, 
we tried to supply the ``jitter'' with unbiased, flat-fielded,
sky-subtracted frames while disabling the sky calculation option of
the recipe. Since this method reproduced the same fluctuation effects  
in the co-added frames, we decided to use the pipeline-reduced frames.\\

PSF-fitting photometry was carried out for all 
filters using the DAOPHOT II/ALLSTAR procedures under the ESO MIDAS
reduction package. Finally the magnitudes were calibrated using the
mean atmospheric extinction coefficients supplied by ESO, and three
standard stars for determining the zero points.  
We used the color equations provided by ESO and 
checked that the color terms were small. \\

We compared our photometry with those provided by the 2MASS point
source catalogue \citep{2mass} using a selection of 15 stars 
which appeared as single in our images 
and were brighter than 15.5 mag in {\it Ks}. Our photometry
agrees well with 2MASS for $J$ and $H$ filters, although is slightly
fainter; the mean differences being m(2MASS)\,--\,m(ISAAC)\,=\,$-0.05$ mag
in $J$ and $-0.07$ mag in $H$. The disagreement is more significant for
the {\it Ks} band, where the mean difference amounts to $-0.44$
mag. The comparison of the filter profiles between ISAAC and 2MASS
systems showed no significant difference. This discrepancy can
therefore be explained by the fact that the {\it Ks} observations were
carried out on a different night, as mentioned above.  During that
night the sky transparency was lower and the seeing poorer. 
We therefore decided to bootstrap our data with the 2MASS
photometry by making the appropriate offset to our measured {\it Ks}
magnitudes. Our final astrometry was based on 11 stars of the field,
the accurate positions of which were determined in our previous {\it HST}
observations (Paper I). \\

\section{Results and discussion}

\begin{figure*}
\begin{center}
\resizebox{17cm}{!}{\includegraphics{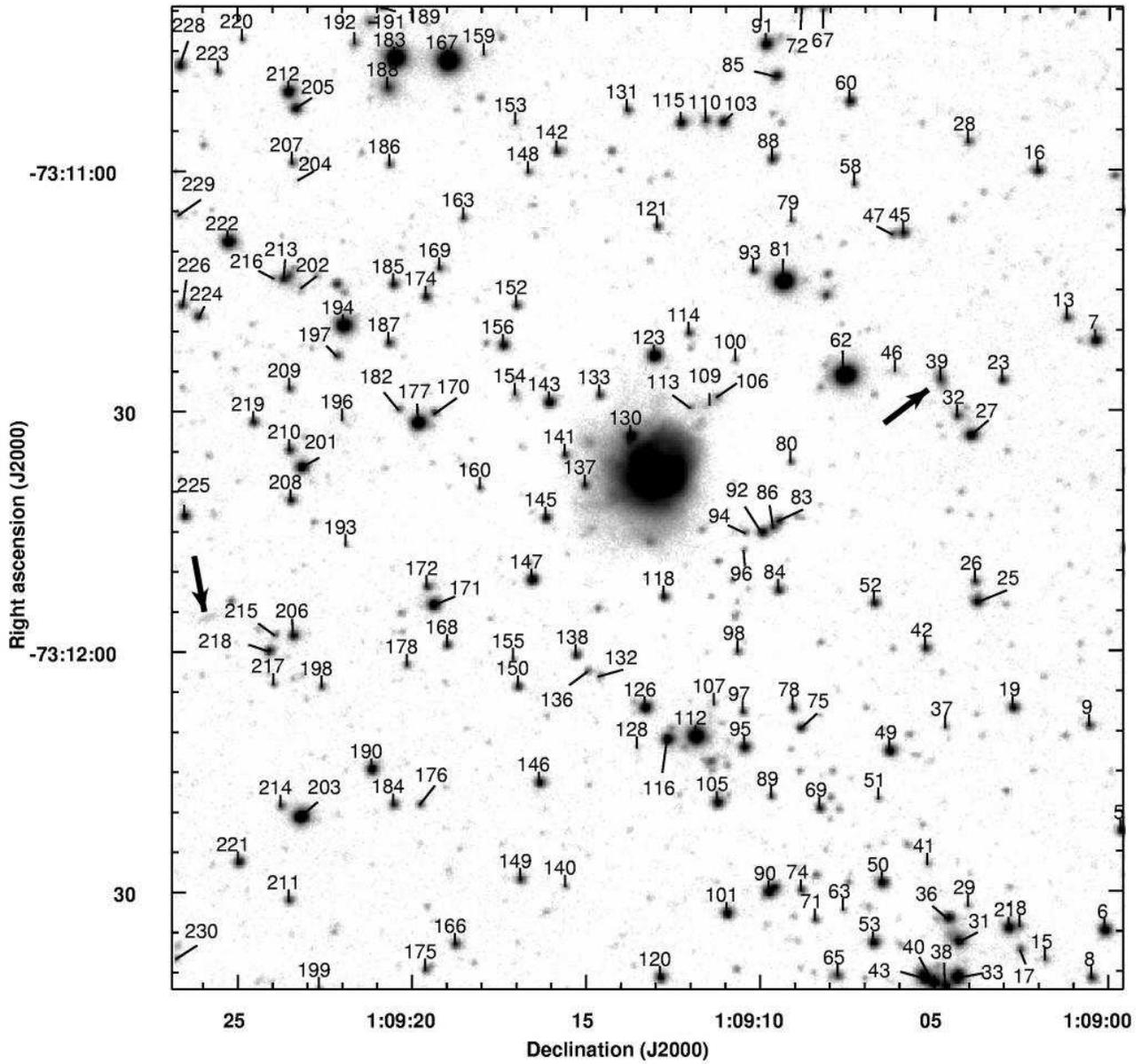}}
\caption{The SMC N81 region and its surrounding field as seen through 
the $J$ filter. The image, taken with VLT/ISAAC, results from the
coaddition of 10 basic exposures.  The field size is
2\min\,\,\x\,2\min\, corresponding to 38 pc $\times\,$ 38 pc.  The
arrows indicate the candidate background galaxies (Section
\ref{partie_galaxies}).  North is up and
east to the left.
\label{chart} 
} 
\end{center}
\end{figure*}

\begin{figure*}
\begin{center}
\hspace*{1.7cm}
\resizebox{10.2cm}{!}{\includegraphics{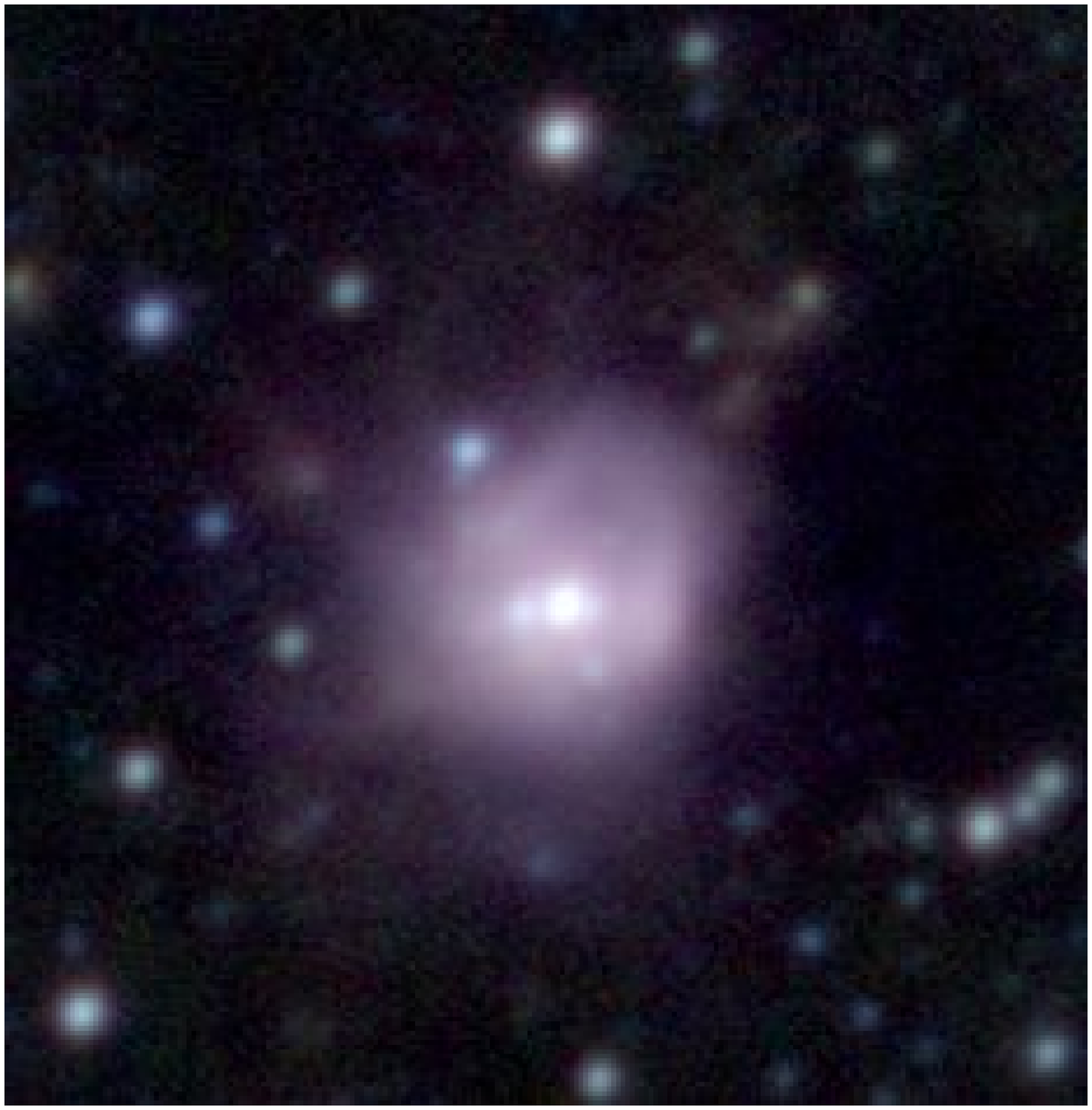}}
\resizebox{12.6cm}{!}{\includegraphics{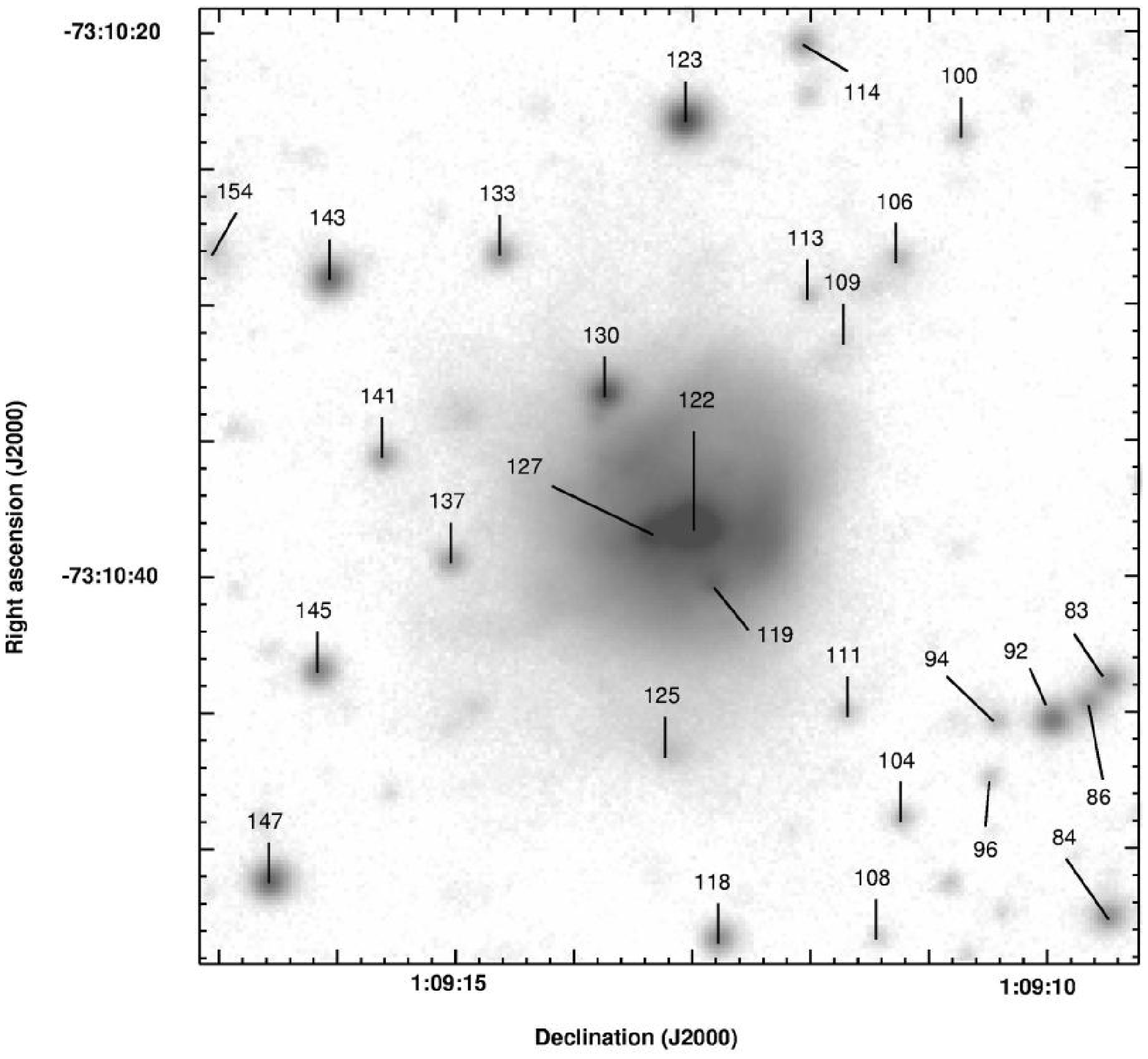}}
\end{center}
\caption{The SMC \h2 region N81 and its immediate field. {\bf a)} 
composite {\it JHKs} color image,  {\bf b)} as in (a) with stars identified.  
\label{close-up}} 
\end{figure*}

A typical final image obtained with ISAAC is presented in
Fig.\,\ref{chart}, while a close-up view of the \h2\, region is
displayed in Fig.\,\ref{close-up}.  The two brightest stars of N81
lying towards the central area of N81, detected by {\it HST} (Paper I)
and separated by 0\frac.27, are not resolved on this image.  The cross
references for common detections in both ISAAC and {\it HST} are
presented in Table\,\ref{corres}. Even though we used broad band
filters for our imaging in order to sample the properties of the
stellar continuum emission, we do detect faint levels of diffuse
near infrared light from the central 10\frac\, of the region. We note that
the spatial extent of this emission is similar in size to that in
H$\alpha$ observed by {\it HST} (see Fig. 1 in Paper I). 
This is mainly due to nebular infrared emission lines
(i.e. Br$\gamma$) with contribution from free-free, two-photon, and 
probably dust emission.\\

A total of 519 stars are detected in the field with magnitudes ranging
from 12.5 to 20.7 in the {\it Ks} band. The internal photometric
errors derived from DAOPHOT are very small ranging from 0.005 to 0.06
mag for {\it Ks} from 14 to 19 mag respectively. However, the true
accuracy is smaller, a cutoff lower limit of {\it Ks}\,=\,19 mag
allowing an estimated accuracy of 0.2 mag on the faintest stars.  \\

\subsection{Color-magnitude and color-color diagrams}

\begin{figure*}
\resizebox{\hsize}{!}{
\includegraphics{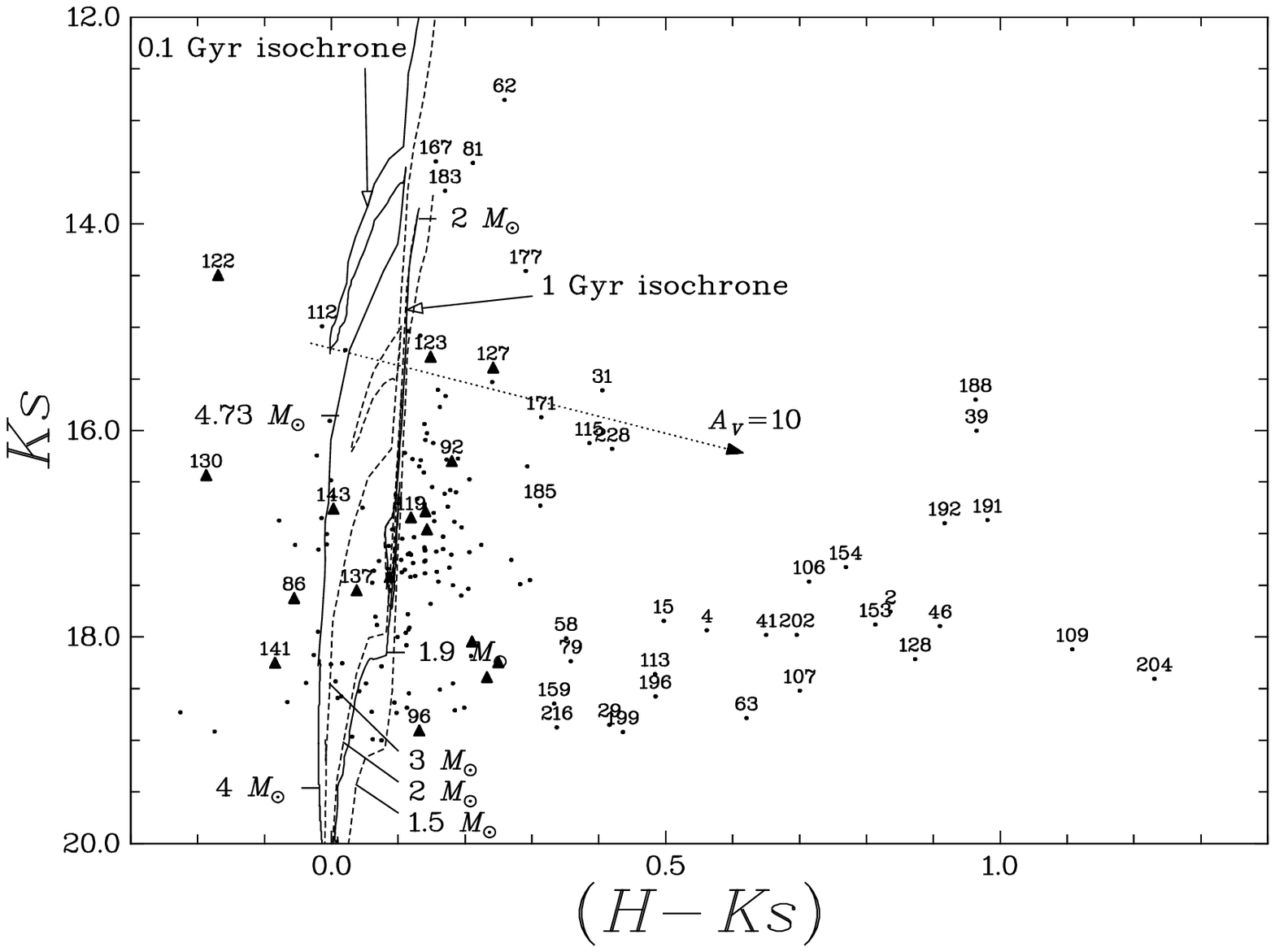}
}
\caption{
Color-magnitude, {\it Ks} versus {\it H\,--\,Ks}, diagram for the
observed stars towards SMC N81. The solid curves show isochrones
predicted by the Geneva group \citep{lejeune01}. The left curve
represents an age of 0.1 Gyr and the right one 1 Gyr. Two mass
boundaries are marked on each isochrone, 4 and 4.73\,\sm\, on the left 
isochrone and 1.9 and 2\,\sm\, on the right one.
 The dashed curves
are evolutionary tracks for masses 1.5, 2, and 3\,\sm.
The reddening track is plotted with
a dotted line and extends to $A_{V}$\,=\,10 mag.
Note that an $A_{V}$\,=\,10 mag, much too high for the area of 
SMC N81 (see text) would be needed for interstellar reddening to 
create the different color 
properties of the third stellar population we identified.  
The numbers refer to the
stellar identifications presented in Fig. \ref{chart} and 
Table \ref{corres}, and triangles represent stars that are 
also identified in paper I.
\label{col-mag} 
}

\end{figure*}

\begin{figure*}
\resizebox{\hsize}{!}{
\includegraphics{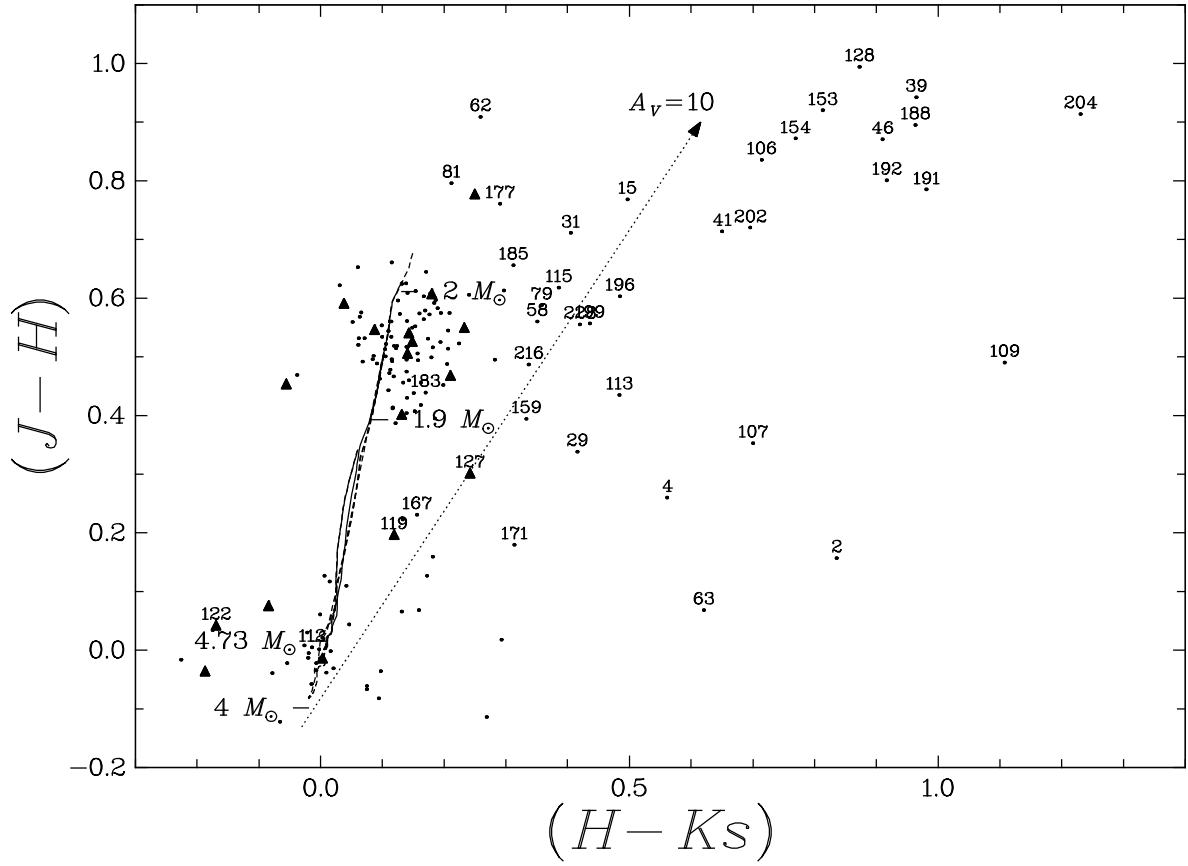}
}
\caption{
Color-color, {\it J\,--\,H} versus {\it H\,--\,Ks}, diagram for the
observed SMC N81 stars. See the Fig.\,\ref{col-mag} caption for
explanations regarding the model fits. Note that here the 0.1 Gyr
isochrone is indicated as a solid
line and the 1 Gyr curve as a dashed one. Triangles
represent stars that are also identified in paper I.
\label{col-col} 
} 
\end{figure*}

In Fig.\,\ref{col-mag} we present the {\it Ks} versus {\it H\,--\,Ks}
diagram of the observed stars in the field of N81.  All sources
brighter than 19th mag in {\it Ks} are present, and those brighter
than {\it Ks}\,=\,15 or redder than {\it H\,--\,Ks}\,=\,0.3 mag are
labelled according to Fig.\,\ref{chart}.  The color-magnitude diagram
shows the presence of a main stellar population
vertically distributed along {\it H\,--\,Ks}\,\ab\,0.1 mag and a 
second group with significantly redder colors. There may also exist
a third smaller population centered at {\it H\,--\,Ks}\,\ab\,0.0, but 
we are not sure because of the photometric uncertainties.
The populations  are also visible on the color-color diagram
(Fig.\,\ref{col-col}). \\

In order to explain the color-magnitude and color-color diagrams, we
compared them with the predictions of the Geneva evolutionary models 
\citep{lejeune01}. To our surprise, we could not find any massive stars
isochrones fitting both diagrams coherently. The best results were
found for masses around 2 and 5\,\sm. Our analysis suggests that 
the bulk of the stars, distributed along {\it
H\,--\,Ks}\,\ab\,0.1 mag, belongs to an evolved population of mass 
\ab\,2\,\sm\, and age 1 Gyr. Moreover, the {\it
H\,--\,Ks}\,\ab\,0.0 population is simulated by \ab\,5\,\sm\, 
model stars of age
1 Gyr.   However, the discrimination between these two 
populations is difficult 
as it is very sensitive also to the interstellar
extinction. A small differential extinction is sufficient to move stars
from one  population to the next. \\

\begin{figure*}
\begin{center}
\resizebox{14cm}{!}{
\includegraphics{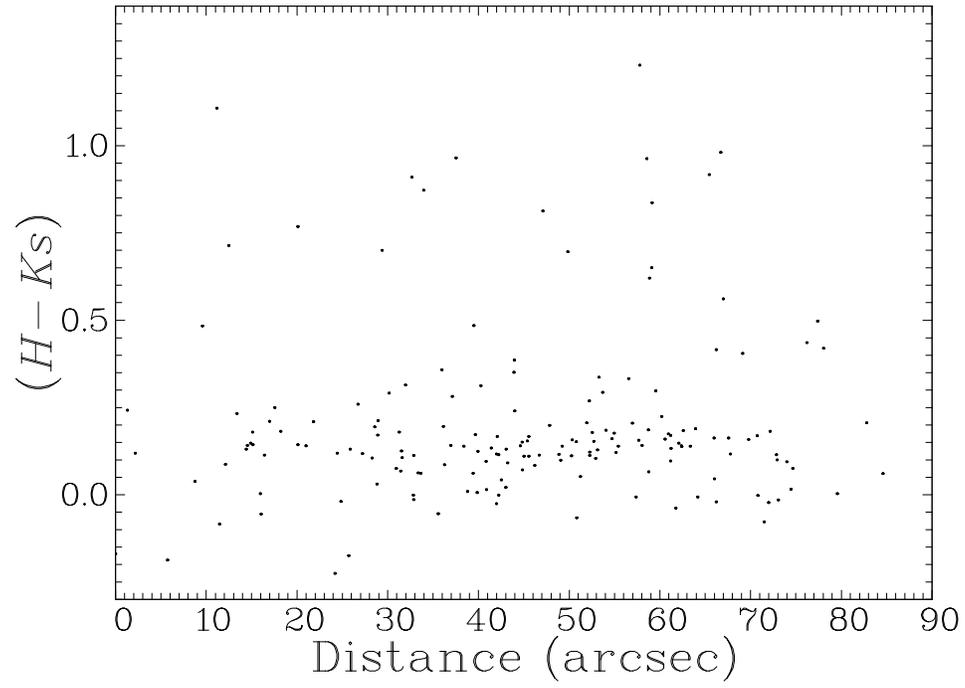}
}
\caption{
Variation of color index {\it H\,--\,Ks}  as a function of the angular 
distance from the center of N81.
\label{spatial-color} 
} 
\end{center}
\end{figure*}

However, several previous works suggest that the interstellar extinction 
towards SMC N81 is rather low. This was also confirmed by the first high 
resolution extinction map of N81, derived from our {\it
HST} observations (Paper I).  Those observations showed that  the
extinction varies across N81, but the higher values do not exceed 
$A_{V}$\,\ab\,1.3 mag, while the mean value is  $A_{V}$\,\ab\,0.40
mag, if the interstellar reddening law is used.  Although this
extinction is derived from the
\ha/\hb\, Balmer decrement which could  be biased towards less reddened
regions, it has also been confirmed by measurements in the near
infrared.  In fact
\citet{israel88} found 0 $<$ $A_{V} \leq 0.6 $ from Br$\gamma$/\hb\,
and Br$\alpha$/\hb\, ratios.  We note also that the ``nebular''
Brackett/Balmer suffers, albeit in a lesser degree, 
from the same scattering
out/in the beam versus absorption phenomenon as the Balmer/Balmer.
However, a small Brackett/Balmer ratio does not necessarily imply a high
extinction!  The low extinction towards N81 is also supported by the
fact that the color index {\it H\,--\,Ks} of the stars in our field
does not display any trend/correlation with the projected radial
distance from N81 (Fig.\,\ref{spatial-color}).  If N81 was
responsible for the extreme reddening of some sources, one should
expect them to lie in its vicinity. On the contrary, this plot shows
that the reddest sources are randomly spread on the field, restricting
the influence of N81 on the color of its neighbours to a radius of
5\frac\, at most.  There is no evidence of segregation of red sources
towards a particular area in our observed field. \\

Another argument against the presence of important 
dust concentration towards N81 comes from the fact that
no important molecular cloud has been detected in the region.  
\citet{israel93}  detected $^{12}$CO\,(1\,-\,0) emission at two
points towards N81 using SEST, even though the beamsize was 43\frac\,
(\ab\,13\,pc), almost \ab\,4 times the size of the \h2\, region.  The
main CO emitting position, lying at the southwestern near periphery of
the \h2 region, has a beam brightness temperature of 375\,mK, a
linewidth of 2.6\,km\,\sec, an LSR velocity of 152 km\,\sec, and an
intensity of 1.0 K\,km\,\sec.  We tried to estimate the corresponding
visual extinction from these molecular observations.  Assuming
molecular hydrogen densities of 1000-3000 cm$^{-3}$, cloud
temperatures of 10-40 K, and a CO/H$_{2}$ abundance ratio of
10$^{-5}$, we get CO column densities of 7--8\,\x\,10$^{14}$
cm$^{-2}$, corresponding to a visual extinction of \ab\,0.08 mag, if
the beam is filled. A filling factor of 10\% leads to column densities
a factor of 10 larger and a visual extinction still smaller than 1
mag. Compared to other neighboring \h2 regions, the main CO position
towards N81 is brighter than those detected towards N76, N78, and N80,
but is weaker than that associated with N84 and more especially
N88. \\

The disperse group of red stars we identified is  not 
a result of extinction but they must represent
 a mix of evolved stars of the SMC lying along the line of sight
of N81. The brightest star of the population,
\#188, has a bolometric magnitude of \ab\,18.5 mag which at the
distance of the SMC translates to 400--500 $L_{\odot}$.  Consequently
it very likely marks the tip of the red giant branch of the stellar
population in the field of N81, even though we would need to obtain
its spectrum to confirm this interpretation.  Two other sources, \#109
and \#204, stand out with {\it H\,--\,Ks} $>$ 1 and {\it Ks} $\geq$ 18
mag and luminosities of \ab\,50 $L_{\odot}$. If this population
actually consists of evolved red giants, it should be much older than
the two other populations.  Alternatively, these reddest colors may
be due to circumstellar emission in young stellar objects, as found in
the $\rho$\,Oph molecular cloud \citep{greene95}. However, 
since the extinction is very low towards SMC N81 and no
important molecular clouds have been detected, this explanation seems 
less plausible.   \\

\begin{table}[htb]
\setcounter{table}{1}
\caption[]{Field stars comparison }
\label{field}
\begin{tabular}[h]{r r r r }
\hline
Nebula & 2\min.5 & 5\min\, & 10\min\,  \\
\hline
SMC N81 &  56  & 359 & 1197 \\
N88 &  28  & 132 &  484 \\
N90 &  39  & 102 &  369 \\
N66 & 192  & 805 & 2883 \\
N70 & 142  & 581 & 2352 \\
\hline
\end{tabular}
\end{table}

In order to determine if our sample is contaminated by the
background/foreground sources, we compared a few SMC fields through
the 2MASS catalog. Using Aladin, we searched for sources around 5
nebulae (N81, N88, N90, N66, N70) for various surrounding areas
(annuli with radii 2.5, 5, and 10 arc-minutes around the core
objects), and compared the number of detections in each case
(Table\,\ref{field}). The smallest annulus corresponds well with the
typical size of a globular cluster in the SMC \citep{hodge85} and is
roughly twice our ISAAC field. The \h2 regions N81, N88, and N90 lie
in  Shapley's wing, while N66 and N70 belong to the denser main body
of the SMC and are much larger. In fact the stellar environment of N81
should preferably be compared with that of N88, since both are very
young HEBs produced by newborn massive stars. Although N90 lies
towards  Shapley's wing, it is more extended, less excited and does
not belong to the HEB class; it is probably older than N81 and
N88A. Table\,\ref{field} shows that the N81 field is richer than those
of N88 and even N90, while we expected similar detection numbers based
on their comparable evolutionary stages. Since N88 is associated with
a large molecular cloud \citep{rubio96} and is affected by a
significantly stronger extinction, we argue that the higher number of
stars detected towards N81 is due to our ability to probe deeper into
the SMC in that region.  We are in fact sampling all the stars
belonging to different star formation events in that direction. \\

Previous work has shown the presence of large complexes of blue stars
in the inter-Cloud region. \citet{grondin92} found associations as
young as 16 Myr with masses in the 1.5--12\,\sm\, range, while
\citet{irwin90} found older blue stars of age 
\ab\,0.1 Gyr. However, these studies concern inter-Cloud areas 
with Right Ascension $\alpha > $2\,h, that is towards the Bridge
central parts and significantly east of the N81 region.  Tidal models
predict that the Bridge was pulled from the SMC during a close
encounter between the two clouds 0.2 Gyr ago \citep{gardiner96}, and
the age of many of the stellar concentrations (10--25 Myr) indicates
that the Bridge is a star forming region.  The star populations
identified in the present study are older than those previously
detected in the Bridge region. They may have formed as a result of an
earlier tidal interactions 1.5 Gyr ago \citep{murai80}.  This is
consistent with \citet{kunkel97}'s result who found a population of
intermediate age carbon stars, few Gyrs in age, scattered throughout
the Bridge region. Model simulations suggest that the carbon stars are
a population of the SMC halo stars displaced into the inter-Cloud
region by tidal forces \citep{gardiner96}. \\

Finally,  the main sequence massive stars identified using {\it HST}
observations (Papers I and II) are also present on the
color-magnitude and color-color diagrams. Shown with filled triangles
in Figs.\,\ref{col-mag}\,\&\,\ref{col-col}, they are mostly the bluest
stars of the sample, while some  are affected by local
extinction. \\

\subsection{Observing background galaxies through SMC}
\label{partie_galaxies}

We have detected two non-stellar objects on the near infrared
images. The source labelled \#39 (Fig.\,\ref{chart}) is a rather
bright, very reddened object with {\it Ks}\,\ab\,16 and {\it H -- Ks}
$>$ 0.9 mag.  It has an elongated shape with an apparent size of
3\frac.5 $\times$ 1\frac.5, and is situated at the J2000 coordinates
$\alpha$ = 01:09:04.85, $\delta$ = --73:11:26.3.  It is very likely
that this source be a background galaxy, but no extragalactic source
at that location has been catalogued.  The possibility of the object
being a circumstellar disk can be ruled out because of its size which
would be 2.2 pc, about 10 times larger than the circumstellar disks
found for example in the Orion nebula \citep{Brandner00}. Another
possibility may be a blending of three aligned red stars.  There is at
least one additional ``background galaxy'' visible diagonally opposite
to \#39 from the field centre, at coordinates $\alpha$ = 01:09:25.8,
$\delta$ = --73:11:55.9.  These detections further underline the
``transparency'' of the SMC in that direction.  It would be
interesting to obtain spectra of these galaxy candidates in order to
elucidate their nature.

\section{Conclusions}

The young massive star forming region N81, lying in Shapley's wing
where the SMC bridges to the LMC, is surrounded by three main
populations of lower mass stars.  These are evolved stars of ages at
least 0.1 to 1.0 Gyr and initial masses of no more than \ab\,2 to
5\,\sm. It appears that the newborn massive stars exciting the \h2
region N81 are not associated with these populations which presumably
represent various star formation events in the SMC along the line of
sight towards N81.  Of course we cannot firmly exclude the
co-spatiality of these populations with N81. However, the relative
``transparency'' of the SMC towards N81, as supported by this work,
favors our interpretation, since the SMC is known to have an overall
complex structure with several overlapping neutral hydrogen layers
\citep{mcgee81}. There may exist a few runaway massive stars
contaminating the low mass population, but the absence of any \h2
region in the field, apart from N81, is in line with the lack of high
mass stars. Anyhow, the number of Magellanic Cloud main sequence stars
counts for a small fraction of the total number of stars in the Wing
region \citep{irwin90}.

\begin{table*}[htb]
\setcounter{table}{0}
\caption[]{Photometry of common {\it HST}/ISAAC stars }
\label{corres}
\begin{tabular}[h]{c c c c c c c c p{.3\linewidth}}
\hline
Star &  $\alpha$ & $\delta$ & $J$  & $H$ & $K$ & $I$  &{\it HST} number \\
     &  (2000.0) &  (2000.0) &  (mag)           &    (mag) & (mag)& (mag) & (Paper I)  \\
\hline
\input{ms3992tab.tex}
\hline
\end{tabular}
\end{table*}

\bibliographystyle{aa}
\bibliography{ms3992}
\end{document}

%% file: ms3992tab.tex
122 & 1:09:13.05& -73:11:38.2 & 14.37 & 14.33 & 14.50 & 14.51 &1 + 2 \\
127 & 1:09:13.34& -73:11:38.4 & 15.94 & 15.63 & 15.39 & 16.10 &3 \\
119 & 1:09:12.83& -73:11:40.2 & 17.16 & 16.96 & 16.84 & 17.57 &8 \\
130 & 1:09:13.74& -73:11:33.3 & 16.21 & 16.25 & 16.43 & 16.11 &11 \\
143 & 1:09:16.07& -73:11:29.1 & 16.75 & 16.76 & 16.76 & 16.67 &13 \\
141 & 1:09:15.62& -73:11:35.6 & 18.24 & 18.16 & 18.25 & 18.14 &14 \\
147 & 1:09:16.58& -73:11:51.2 & 16.63 & 16.17 & 16.03 & 17.21 &15 \\
111 & 1:09:11.70& -73:11:45.0 & 19.34 & 19.18 & 19.10 & 19.07 &17 \\
125 & 1:09:13.20& -73:11:46.5 & 19.48 & 19.43 & 19.33 & 19.13 &18 \\
118 & 1:09:12.78& -73:11:53.4 & 17.64 & 17.10 & 16.96 & 18.49 &19 \\
94 & 1:09:10.43& -73:11:45.4 & 19.17 & 18.62 & 18.39 & 17.91 &20 \\
92 & 1:09:09.96& -73:11:45.3 & 17.09 & 16.48 & 16.30 & 18.85 &21 \\
86 & 1:09:09.66& -73:11:44.7 & 18.02 & 17.57 & 17.62 & 18.73 &22 \\
137 & 1:09:15.05& -73:11:39.5 & 18.17 & 17.58 & 17.55 & 19.11 &23 \\
145 & 1:09:16.17& -73:11:43.5 & 17.43 & 16.93 & 16.79 & 18.10 &24 \\
133 & 1:09:14.63& -73:11:28.2 & 18.05 & 17.50 & 17.42 & 18.68 &26 \\
123 & 1:09:13.07& -73:11:23.3 & 15.96 & 15.44 & 15.29 & 16.82 &27 \\
95 & 1:09:10.48& -73:12:12.1 & 17.03 & 17.06 & 17.11 & 19.66 &29 \\
134 & 1:09:14.67& -73:11:54.8 & 18.92 & 18.87 & 19.64 & 19.04 &30 \\
96 & 1:09:10.49& -73:11:47.4 & 19.44 & 19.03 & 18.90 & 19.78 &33 \\
80 & 1:09:09.15& -73:11:36.5 & 18.72 & 18.25 & 18.04 & 20.16 &35 \\
203 & 1:09:23.22& -73:12:20.7 & 15.44 & 15.22 & 15.08 & 15.84 &-- \\
222 & 1:09:25.29& -73:11:08.9 & 15.83 & 15.77 & 15.61 & 15.73 &-- \\
201 & 1:09:23.19& -73:11:37.1 & 16.38 & 15.77 & 15.53 & 17.36 &-- \\
190 & 1:09:21.17& -73:12:14.8 & 16.68 & 16.27 & 16.12 & 17.39 &-- \\
212 & 1:09:23.55& -73:10:50.3 & 16.35 & 15.93 & 15.77 & 16.97 &-- \\
194 & 1:09:21.98& -73:11:19.4 & 15.21 & 15.24 & 15.22 & 15.03 &-- \\
171 & 1:09:19.41& -73:11:54.3 & 16.37 & 16.19 & 15.87 & 16.47 &-- \\
177 & 1:09:19.84& -73:11:31.6 & 15.51 & 14.75 & 14.46 & 16.66 &-- \\
183 & 1:09:20.51& -73:10:46.0 & 14.29 & 13.85 & 13.68 & 14.96 &-- \\
126 & 1:09:13.32& -73:12:07.2 & 16.48 & 15.84 & 15.67 & 17.49 &-- \\
116 & 1:09:12.68& -73:12:11.1 & 16.47 & 16.48 & 16.48 & 16.35 &-- \\
112 & 1:09:11.87& -73:12:10.8 & 14.98 & 14.98 & 14.99 & 14.87 &-- \\
90 & 1:09:09.76& -73:12:30.2 & 16.66 & 16.64 & 16.35 & 16.40 &-- \\
49 & 1:09:06.30& -73:12:12.6 & 16.77 & 16.33 & 16.22 & 17.57 &-- \\
43 & 1:09:05.31& -73:12:40.8 & 15.91 & 15.91 & 15.91 & 15.97 &-- \\
40 & 1:09:05.03& -73:12:41.5 & 16.25 & 16.22 & 16.24 & 16.19 &-- \\
37 & 1:09:04.69& -73:12:41.9 & 19.34 & 18.88 & 18.69 & 16.75 &-- \\
36 & 1:09:04.62& -73:12:33.5 & 16.84 & 16.79 & 16.75 & 16.86 &-- \\
33 & 1:09:04.36& -73:12:40.9 & 15.81 & 15.15 & 15.04 & 16.90 &-- \\
21 & 1:09:02.87& -73:12:34.8 & 16.75 & 16.79 & 16.87 & 16.69 &-- \\